\documentclass[aps, prd, twocolumn, lengthcheck, superscriptaddress, showpacs, letterpaper, nofootinbib]{revtex4-1}

\usepackage{cleveref}
\usepackage{times}
\usepackage{amsmath}
\usepackage{graphicx}
\DeclareGraphicsExtensions{.pdf,.eps,.png,.jpg}
\usepackage{acronym}
\usepackage{color}
\usepackage[caption=false]{subfig}
\usepackage{tabularx}
\usepackage{longtable}
\usepackage{multirow}
\usepackage{dcolumn}
\usepackage{soul}
\usepackage{float}
\usepackage{hyperref}

\allowdisplaybreaks

\def\be{\begin{eqnarray}}
\def\ee{\end{eqnarray}}

\def\roughly#1{\mathrel{\raise.3ex\hbox{$#1$\kern-.75em%
\lower1ex\hbox{$\sim$}}}}

\def\gsim{\roughly>}

\begin{document}

\title{A Stiffer EoS for Compact Stars in Effective Field Theory}
\author{Kyungmin \surname{Kim}}
\affiliation{Department of Physics, Hanyang University, Seoul 133-791, Korea}

\author{T.T.S. \surname{Kuo}}
\affiliation{Department of Physics and Astronomy, State University of New York, Stony Brook, New York 11794, USA}

\author{Hyun Kyu \surname{Lee}}
\affiliation{Department of Physics, Hanyang University, Seoul 133-791, Korea}

\author{Jaehyun \surname{Lee}}
\affiliation{Department of Physics, Hanyang University, Seoul 133-791, Korea}

\author{Mannque \surname{Rho}}
\affiliation{Institut de Physique Th$\acute{e}$orique, CEA Saclay, 91191 Gif-sur-Yvette c$\acute{e}$dex, France}

\date{\today }

\begin{abstract}

We present in this paper a simple and unequivocal prediction on the deformation of the compact star,  that will be measured in gravity waves,  with an EoS obtained in renormailzation-group implemented effective field theory anchored on scale and hidden-local symmetric Lagrangian endowed with topology change . The objective is not to offer a superior or improved EoS for compact stars but to confront with a forthcoming astrophysical observable the given model formulated in what is considered to be consistent with the premise of QCD. The model so obtained is found to satisfactorily describe the observation of a 2-solar mass neutron star \cite{demorest,antoniadis} with a minimum number of parameters.  Specifically the observable we are considering in this paper is the tidal deformability parameter $\lambda$ (equivalently the Love number $k_2$) in gravity waves.  The merit of our prediction is that the prediction can be readily confirmed or falsified by forth-coming  aLIGO and aVirgo gravity-wave observations and provide a valuable guidance for arriving at a better understanding of highly compressed baryonic matter.

\end{abstract}


\maketitle

\section{INTRODUCTION}
The observation of 2-solar mass neutron stars~\cite{demorest,antoniadis} seems to indicate that the equation of state(EoS) for compact stars needs to be sufficiently stiffer to accommodate the mass larger than 1.5-solar mass. This observation led us to try to formulate a field theory framework wherein both low and high density regimes are treated on the same footing. The low-density regime is constrained by experimental data available up to and slightly above nuclear matter density $n_0$ and hence is fairly well controlled theoretically but the high density regime much above $n_0$ is more or less uncharted both experimentally and theoretically, the latter due to the lack of lattice QCD. To cover both regimes in a consistent way in a unified field theoretic approach seems like a tall order, but a first step forward to that goal has been made and this will be reviewed in a long article to appear~\cite{PKLR}.

In this paper, we will pick up the first case studied for the formulation with a confrontation with the observed massive stars~\cite{Dong}, discuss the physical properties of stellar matter with the EoS obtained therein and subject it to an observable different from what's given in \cite{demorest,antoniadis}, which constitutes a prediction that could be tested in gravity waves. The aim here is then to confirm or falsify the strategies taken and assumptions made in \cite{Dong}. The hope is that the result would then help point the directions to be taken in the efforts described in  \cite{PKLR}\cite{Dong}.

Stated briefly without going into the details found in \cite{Dong} which are not needed for this paper, the issue is as follows. When the BR scaling (referred to as ``old-BR") proposed in 1991~\cite{BR} is applied to the neutron-star calculation using realistic NN potentials~\cite{tom}, the mass is estimated to be in the range, $1.2M_{\odot} \sim 1.8 M_{\odot}$,  typically less than the massive stars, $\sim$ 2$M_{\odot}$.  In \cite{Dong}\cite{RhoG}, in the framework that combines scale symmetry and hidden local symmetry of strong interactions for the elementary degrees of freedom, a new density-scaling of the bare parameters of the Lagrangian called ``new-BR/BLPR"  was proposed to incorporate the change in  topology of  the crystal structure of dense skyrmions, skyrmion $\rightarrow$ half-skyrmion~\cite{byp}, which is expected in large $N_c$ QCD. It turns out that this is one option among others that one can take for going to high density with the given Lagrangian~\cite{Dong}. Suppose the threshold density, $n_{1/2}$,  is higher but not much higher than the normal nuclear density $n_0$, then we expect such topology change to have nontrivial effects on nuclear matter at the density $n \geq n_{1/2}$.  It turns out that the effect is drastic, specially in the symmetry energy. As precisely stated in \cite{Dong}, the  change in the symmetry energy observed at the density $n_{1/2}$~\cite{byp}\cite{lr}  can be translated into the parameter changes of the effective Lagrangian, leading to a new scaling in physical quantities. One of the most dramatic effects of the parameter change is the modification of the nuclear tensor forces, thereby drastically affecting the symmetry energy  at high density $n> n_{1/2}$. In a nutshell, what happens is that the contribution to the tensor forces by the exchange of the $\rho$ meson gets strongly suppressed at $n_{1/2}$, thereby increasing the net tensor forces entering into the EoS. In \cite{Dong}, the new BR/BLPR scaling was incorporated into the $V_{lowk}$-implemented nuclear EFT and the mass-radius relation of a compact object of a pure  neutron matter was calculated. It was found that the EoS got stiffer at $n_{1/2}$, giving rise to a star mass as large as $2.4 M_{\odot}$, seemingly consistent with the recently observed high mass neutron stars. In arriving at this result which is consistent with the observables of ~\cite{demorest,antoniadis}, certain assumptions are made that require verification. It is the purpose of this note that some of those assumptions be verified by gravity-wave observations, probing properties different from or additional to those of ~\cite{demorest,antoniadis}.

In what we do below, we make the approach for the compact star more realistic than in \cite{Dong} with electrons, protons and neutrons,  which are believed to be in weak equilibrium, rather than pure neutron matter. Near the surface of star, which is supposed to be  in lower density region, $n < 0.5 \, n_0$, we adopt the equation of state used by  K. Hebeler {\it et al.}~\cite{Astro.J/773/11}.

It is  assumed, in the range of density we are considering, that the energy density of asymmetric nuclear matter ($n_p \neq n_n$ or $x \neq 1/2$) can be described by the conventional form in terms of the symmetry energy factor, $S(n)$, as given by
\be
\epsilon_{nuc} (n,x) =  \epsilon_{nuc} (n,1/2) + n(1-2x)^2 S (n),  \label{enuc}
\ee
where $m_N$ is the mass of nucleon and $x \equiv n_p /n$ is the fraction of proton density.
Then the symmetry energy factor $S(n)$ can be written as
\be
\label{sym_e_fac}
S(n) &=& [\epsilon_{nuc} (n,0) -  \epsilon_{nuc} (n,x=1/2)]/n
\ee
which is equivalent to the difference in the ground-state  energy per nucleon between  the symmetric nuclear matter ($x=1/2$) and  the  neutron matter ($x=0$).  The pressure of nuclear matter is given by $p_{nuc} = n^2\partial (\epsilon(n)/n)/\partial n$.
In this work, we use the corresponding  ground-state  energy  and the symmetry energy factor obtained in \cite{Dong} with the new scaling. The chemical potential difference between proton and neutron  is then given by
\be
\mu_n-\mu_p = 4(1-2x)S(n) \label{munp}.
\ee
 In the  weak equilibrium, the proton fraction, $x$,  is determined  essentially by  the chemical equilibrium condition together with the charge neutrality condition,

Now given an EoS for energy density, $\epsilon$, and pressure, $p$, the radius $R$ and mass $m(R)$ can be determined by integrating the Tolmann-Oppenheimer-Volkoff (TOV) equations \cite{PhysRev/55/364, PhysRev/55/374}.  The equations are integrated up to the radius of the star, $R$, where $p(R) = 0$, and the mass of the star is determined by $m(R)$.  What we are particularly interested in is that the masses and radii, which depend on the equation of state,  are important  in predicting the gravitational waves emitted  from the coalescing binary neutron stars. During the  inspiral period of binary neutron stars,  tidal distortions of neutron stars are expected and the resulting gravitational wave is expected to carry a clean imprint of the equation of states involved~\cite{PRD/81/123016}. The tidal deformability of polytropic EoS, $p = K \epsilon^{1+1/n}$, where $K$ is a pressure constant and $n$ is the polytropic index, were evaluated by Flanagan and Hinderer \cite{PRD/77/021502, ApJ/677/1216} and by others in more detail~\cite{PRD/80/084035, PRD/80/084018}. However, polytropes are known to be a rough approximation to the EoS. In this work, we calculate the mass-radius  and the tidal deformability using the stiffer EoS, which has been recently proposed with the new scaling law (new-BR)~\cite{Dong}.

The equation of state of compact stars with neutrons, protons, electrons and muons in weak equilibrium and charge neutrality condition is discussed and the mass and radius are estimated. We then apply the new stiffer EoS to investigate the tidal deformation of compact stars. The results obtained are clear-cut, parameter-free and could be confirmed or falsified in forthcoming LIGO gravity-wave observations. This confrontation with Nature could give a hint as to whether the novel structure of the EoS predicted by the new scaling law is viable.

We use units in which $c=G=1$ and the notation in which Minkowski metric $\eta_{\mu\nu} = \mathrm{diag}[-1,\ 1,\ 1,\ 1]$.

\section{Equation of state of compact star with neutron, proton, electron and muon}

The asymmetry of neutron and proton numbers at high density, dictated by the chemical potential difference,    inevitably leads to the weak equilibrium configuration with electrons and muons with neutrinos escaped. It can be summarized by the  relation between  chemical potentials: the chemical potential difference between neutron and proton should be the same as the electron chemical potential, $\mu_n-\mu_p = \mu_e = \mu_{\mu}$,
where  the last equality is due to the muon emergence at higher density when  the chemical potential difference from neutron and proton becomes larger than the muon mass.
For  the charge carriers,  protons, electrons and muons with  the local charge neutrality condition, $
n_p = n_e + n_\mu ~,$ one can get the carrier densities at a given density.

The total energy density and pressure  are given by
\be
\epsilon (n,x) = \epsilon_{nuc} + \epsilon_{lep}, ~ ~ ~
p (n,x) = p_{nuc} + p_{lep} ~.
\ee
The energy density, $\epsilon_{lep}$,  and the pressure, $p_{lep}$,  are given by the degenerate  Fermi gas of  leptons (electron and muon) assuming a cold compact star ($T\sim0$).

\begin{figure}[!t]
\subfloat[]{\label{fig3-1}\includegraphics[width=0.45\textwidth]{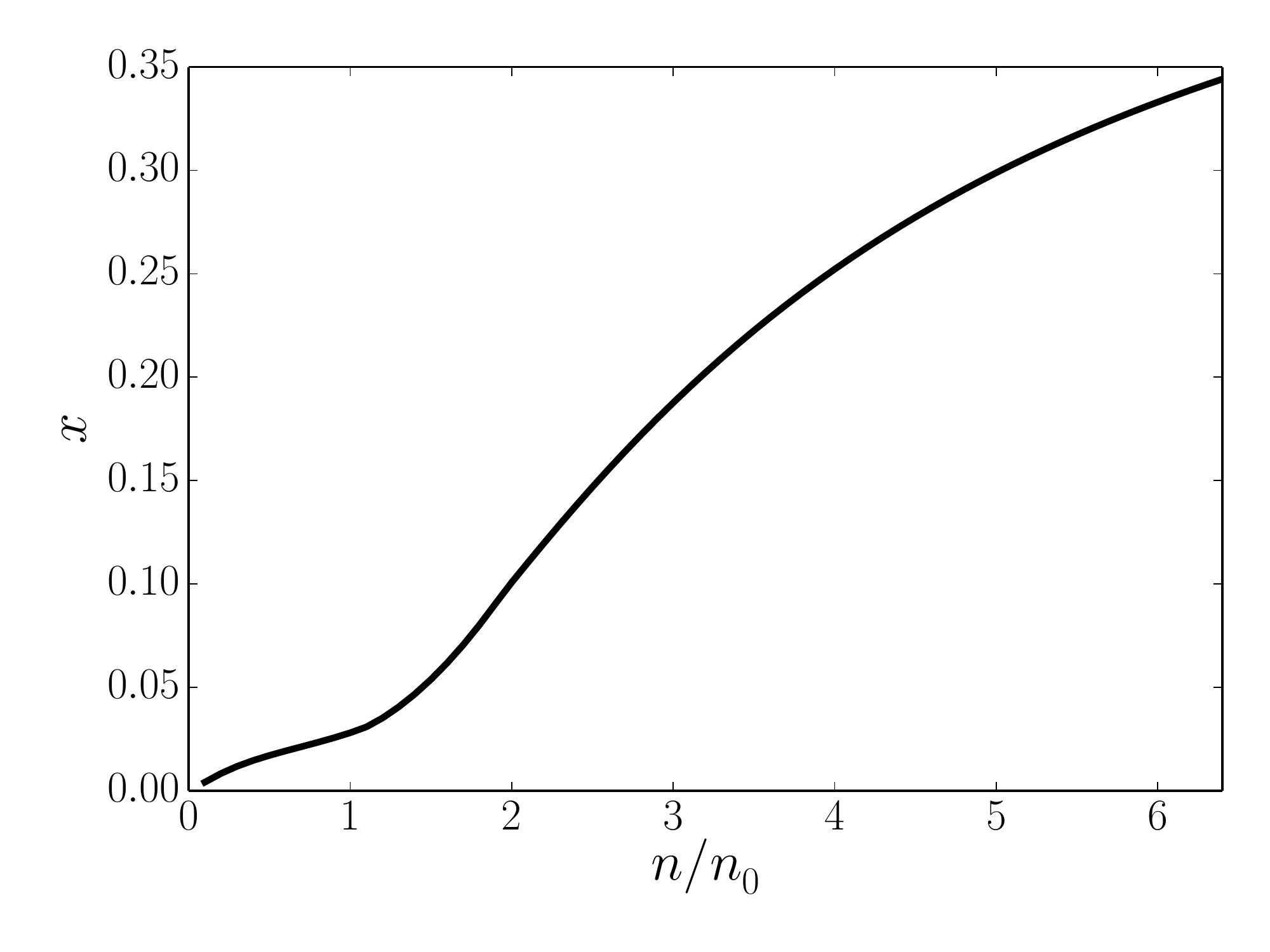}} \\
\subfloat[]{\label{fig3-2} \includegraphics[width=0.45\textwidth]{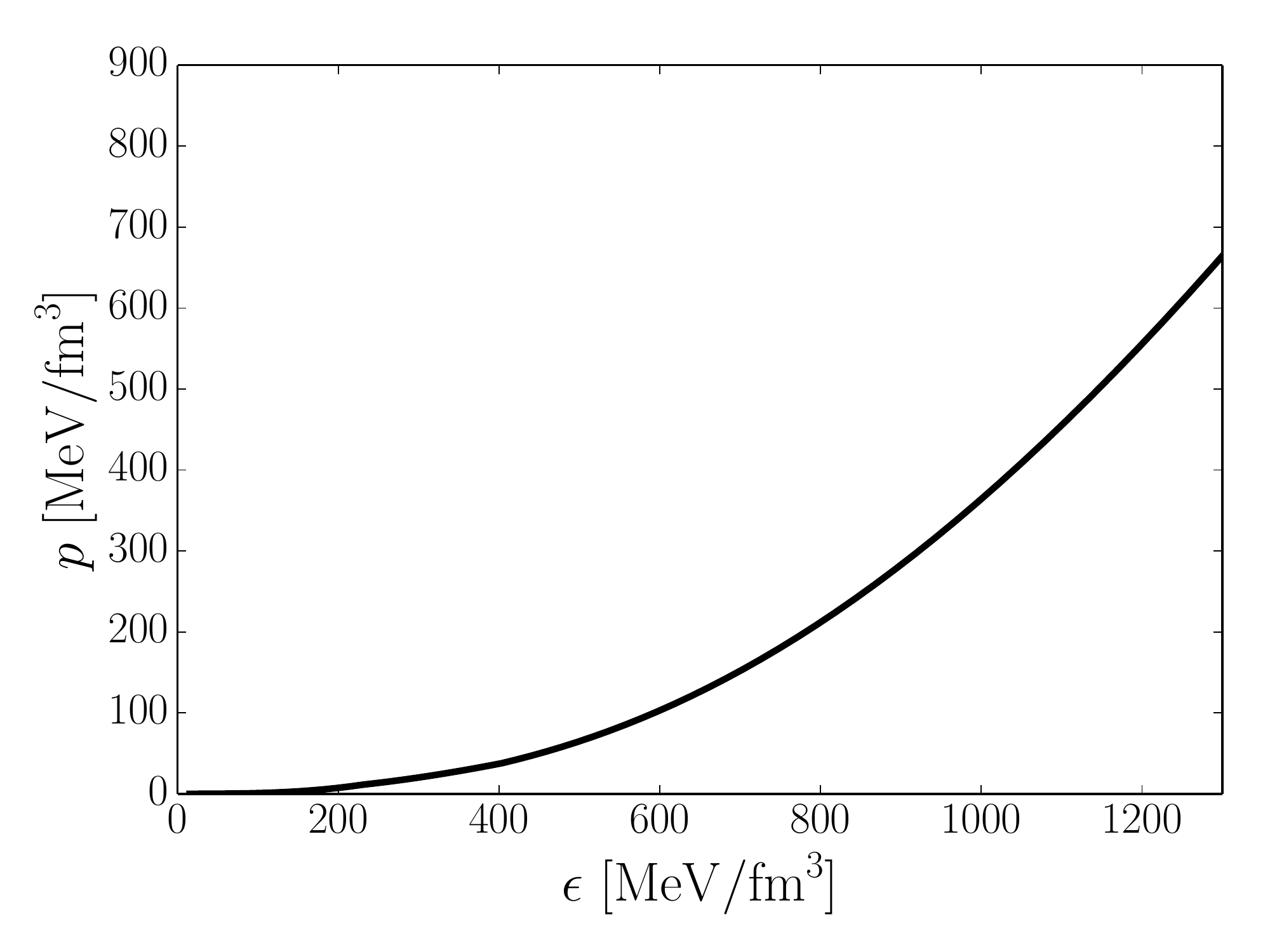}}
\caption{(a) The proton fraction, $x$,  as a function of nuclear density $n$ and (b) energy density($\epsilon$) and pressure($p$),  for an $npe\mu$ configuration. }
\label{fig3}
\end{figure}

The resulting equation of state is shown as a pressure-energy density diagram in Fig. \ref{fig3}. The density dependence of proton fraction, $x$, is shown in Fig. 1(a). One can see that the  proton fraction increases significantly as density increases, because of the increasing chemical potential difference due to the symmetry energy in Eq.(\ref{munp}).
The causality limit, $c_s \leq c$,  constrains the highest density, $n_c$, beyond which  the stiffer EoS used is no longer valid. For the EoS used in  this work,  it  is found to be $n_c \sim 5.7\ n_0$.

For a static and spherically symmetric astrophysical compact star, the metric is given by
\be
ds^2 = -e^{\Phi(r)}dt^2 + e^{\Lambda(r)} dr^2 + r^2 d\theta^2 + r^2\sin^2 \theta d\phi^2,  \label{smetric}
\ee
where $\Lambda$ can be expressed in terms of a radial-dependent  mass parameter, $m(r)$: $e^{\Lambda(r)}  = (1-\frac{2m(r)}{r})^{-1}$.
Assuming a perfect-fluid stellar matter, the relativistic hydrodynamic equilibrium is governed  by  Tolman-Oppenheimer-Volkov(TOV) equation
\be
\frac{dm}{dr} = 4\pi r^2 \epsilon, ~~
\frac{dp}{dr} = - ( \epsilon + p ) \frac{ m + 4\pi r^3 p}{r ( r - 2 m )},\label{TOV2} \\
\frac{d\Phi}{dr} = - \frac{1}{\epsilon +p} \frac{dp}{dr},\label{TOV3}
\ee
where $\epsilon$ and $p$ are energy density and pressure at $r$, respectively. $m(r)$ is the mass enclosed inside the radius $r$. We can calculate the mass of compact star, $M$,  and its radius, $R$,  by integrating the TOV equation up to $p(R)=0$ and get the profiles of $\Phi(r), m(r)$ and $p(r)$.

\begin{figure}[!t]
\includegraphics[width=0.45\textwidth]{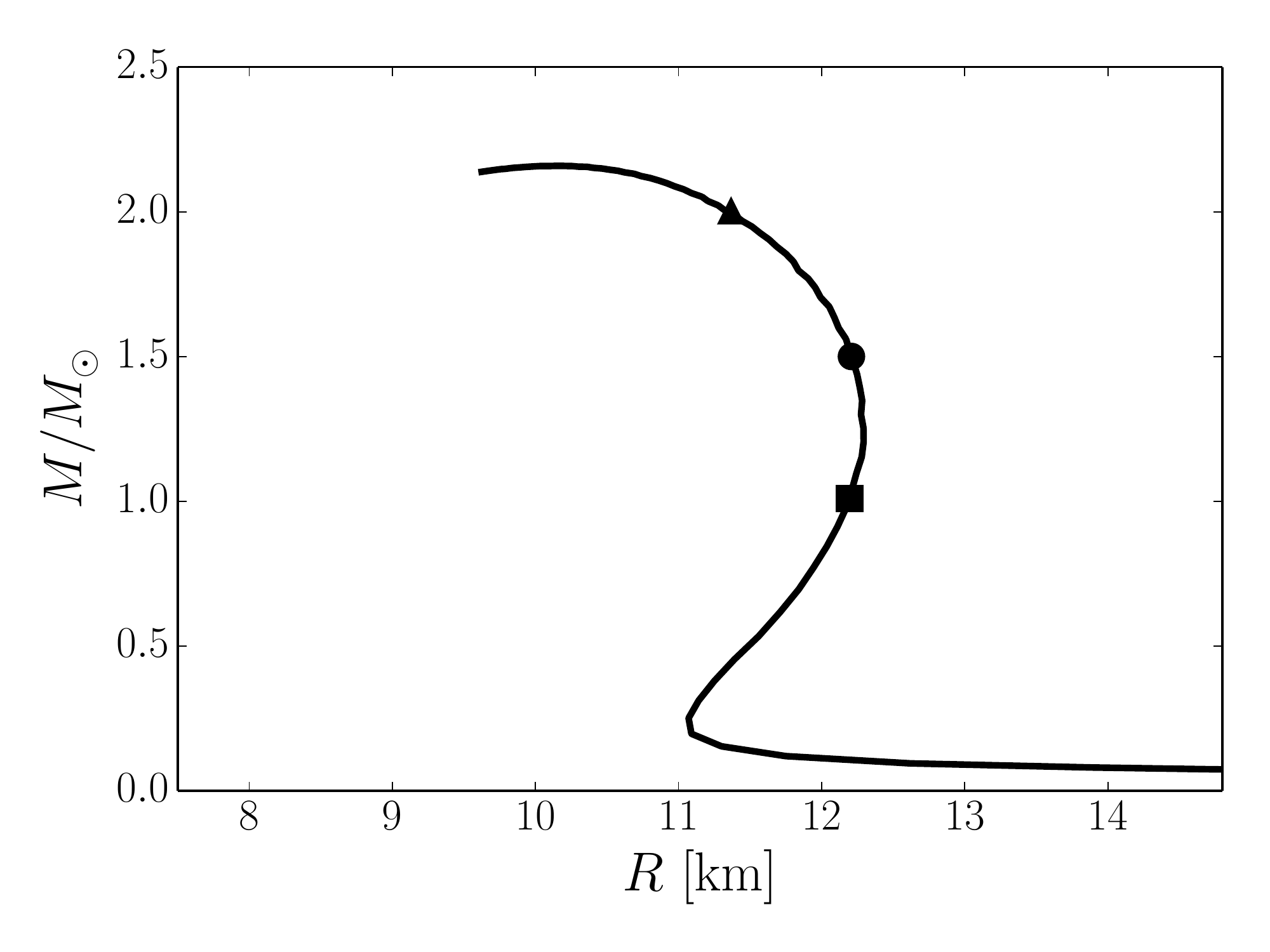}
\caption{Mass($M$)-Radius($R$) curve. The filled-square, filled-circle and filled-triangle correspond to  $M=1.0 M_\odot$ , $M=1.5 M_\odot$ , and $M=2.0 M_\odot$ respectively. }
\label{MR}
\end{figure}

 The EoS of $np$ asymmetric configuration is used to solve TOV equation resulting in the mass-radius curve shown in Figure~\ref{MR}.  For $np$ asymmetric configuration, the possible maximum mass is estimated  to be $M \sim 2.1 M_\odot$ with the radius $R \sim 11$ km, where the central density is about $5.7 \, n_0$. For pure neutron matter~\cite{Dong}, the possible maximum mass is approximately $M \sim 2.4 M_\odot$ with the radius $R \sim 12$ km and $n \sim 4.7 n_0$. In Figure~\ref{MR},  the filled-square, filled-circle and filled-triangle correspond to  $M=1.0 M_\odot$ , $M=1.5 M_\odot$ , and $M=2.0 M_\odot$, respectively. The compactness $C=\frac{M}{R}$ in the range of mass $1.0 - 2 M_{\odot}$ is found to be 0.12 - 0.26 and 0.14 for $1.4 M_\odot$

\section{Tidal Deformation: Deformability parameter(Love number)}
When a nonrotating compact star in a  spherically symmetric configuration is placed in a static external field, it gets deformed by the external  field.  The asymptotic expansion of the  metric at large distances $r$ from the star defines the quadrupole moment, $Q_{ij}$,  and the external tidal field, $\mathcal{E}_{ij}$, as expansion coefficients~\cite{PRD/58/124031} given by
\be
\frac{1 + g_{00}}{2} =  \frac{m}{r} + \frac{3}{2} \frac{Q_{ij}}{r^3} n^i n^j - \frac{1}{2} \mathcal{E}_{ij} r^2 n^i n^j + \cdots ~, \label{quad}
\ee
where $n^i = x^i / r$ and $Q_{ij}$ and $\mathcal{E}_{ij}$ are both symmetric and traceless \cite{PRD/58/124031}.
The deformability parameter $\lambda$ is defined by
\begin{equation}
\label{def}
Q_{ij} = - \lambda \mathcal{E}_{ij} ~,
\end{equation}
which depends on the EoS of baryonic matter and provides the information on how easily the star is deformed. The deformability parameter can be reexpressed by the dimensionless Love number, $k_2$, $\lambda = \frac{2 k_2}{3}R^5 \label{k2}$.

In general,  the  linearized perturbation of the metric caused by an external field is given by \cite{ApJ/149/591},
\begin{equation}
g_{\mu\nu} = g_{\mu\nu}^{(0)} + h_{\mu\nu} ~,\label{hmunu}
\end{equation}
where $g_{\mu\nu}^{(0)}$ is the unperturbed metric in  Eq.(\ref{smetric}).
$h_{\mu\nu}$ is a linearized perturbation, which carries  the information of $Q_{ij}$ and $\mathcal{E}_{ij}$ in Eq.(\ref{quad}). Since we will be considering  the early stage of binary inspiral before the merging stage, the leading order tidal effects with even parity, $l=2$ and $m=0$, are dominant~\cite{PhysRev/108/1063}.  The relevant component for tidal deformation,  $h_{00}$,  for  the static and  even-parity perturbation,  can be written in the following form \cite{ApJ/677/1216}
\be
h_{00} =  -e^{2 \Phi (r)} H(r) Y_{20} (\theta, \phi). \label{hmunu}
\ee
Using  the linearized Einstein equations and the non-vanishing components of the perturbation of stress-energy tensor, we  obtain the differential equation for $H(r)$:
\be
&H''& + \left( \frac{2}{r} + \Phi' - \Lambda' \right) H' + \bigg \{ 2 ( \Phi'' - \Phi'^2 ) - \frac{6}{r^2} e^{2 \Lambda}  \nonumber \\
& &  + \frac{3}{r} \Lambda'  +  \frac{7}{r} \Phi' - 2 \Phi' \Lambda' + \frac{f}{r} ( \Phi' + \Lambda' ) \bigg \} H  = 0 ~, \label{ODE}
\ee
where the prime $'$ denotes the differentiation $d / dr$ and  $f(r) = d\epsilon/dp $.  Using the continuities of $H(r)$ and $H'(r)$ at the boundary, $r=R$,  both for interior and exterior solutions of Eq.(\ref{ODE}), the deformability parameter, $\lambda$,  can be written explicitly~\cite{ApJ/677/1216} in terms of the compactness $C=M/R$ and $y= R H'(R) / H(R)$.

\begin{figure}[!t]
\includegraphics[width=0.45\textwidth]{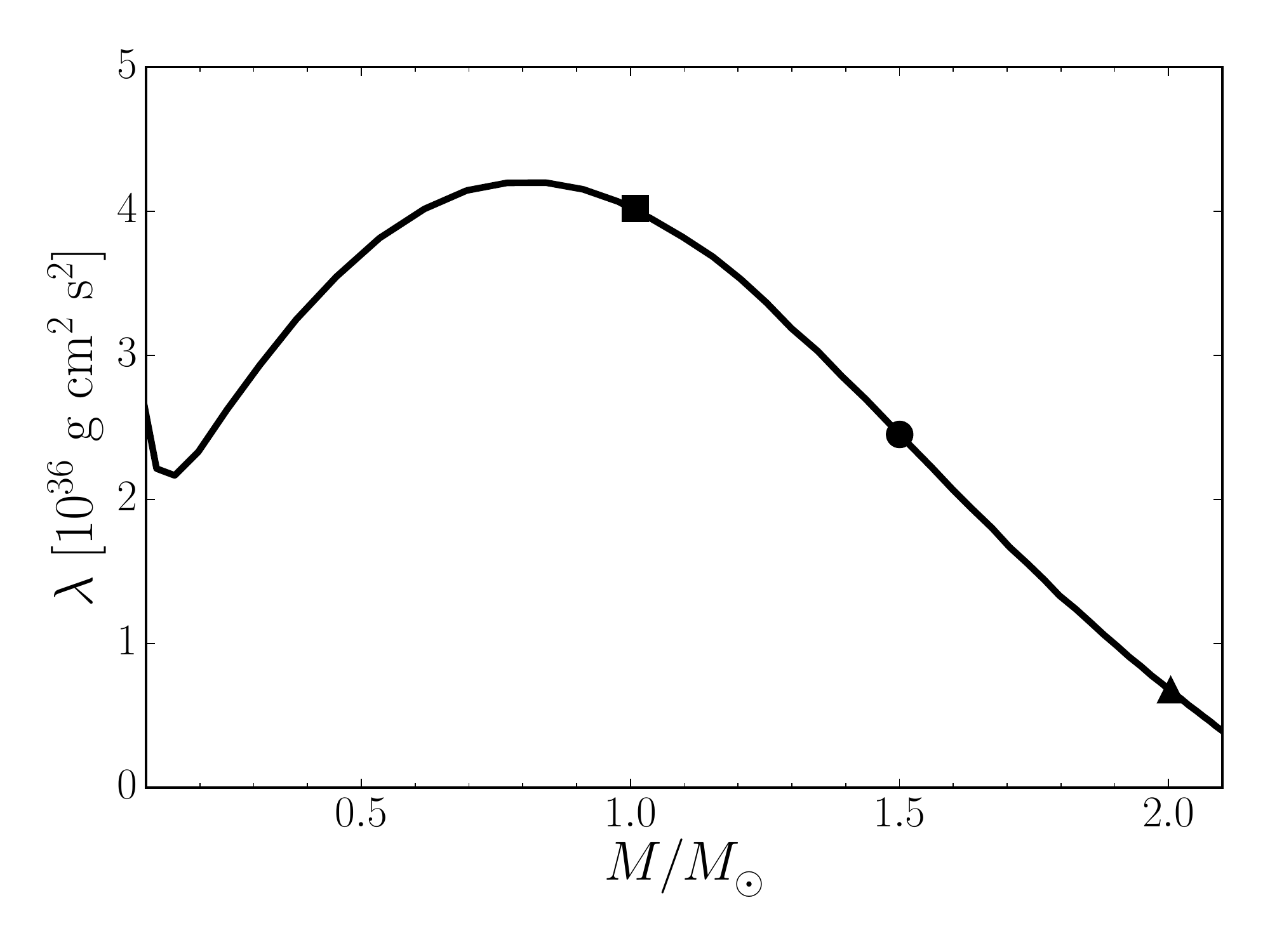}
\caption{The tidal deformability parameter $\lambda$ in the mass range $0.1M_\odot - M_{max}$.}
\label{fig6}
\end{figure}

By  solving TOV equation and Eq.\ (\ref{ODE}) together, we can then calculate  $y$ and the compactness $C$ for the interior solution to obtain  $\lambda$ or Love number,   as shown in Figure~\ref{fig6}.

The deformability parameter  for $1.4M_{\odot}$ is found to be   $2.86 $. It can be compared with those of different EoS's  with only $npe\mu$ matter.   For example, the EoS's of SLy\cite{DH}, AP3\cite{AP3} and  MPA1\cite{MPA}  for the same mass give $\lambda = 1.70, ~2.22$ and $2.79$ respectively~\cite{PRD/81/123016}. On the other hand, the slope of $\lambda$ is found to be stiffer than those above in the mass range $1M_{\odot} ~ - ~ 2 M_{\odot}$.    In the lower mass region around  $\lesssim 1M_{\odot}$, the deformability parameter is found to be relatively higher than those of above EoS's ($\lambda < 3$),  with the maximum value of $ 4.2$   at $0.84M_{\odot}$ .

\section{Summary and further remarks}
 We discussed the physical properties of stellar matter with a new stiffer EoS, which has been proposed recently using a new scaling law (new-BR/ BLPR) in medium caused by topology change at high density~\cite{Dong}, by  extending  Dong {\it et al.}'s work for pure neutron matter to a realistic nuclear matter of  $n$, $p$, $e$ and $\mu$.  The mass- radius and the tidal deformability were calculated.

The calculated maximum mass of compact star is found to be about $2 M_\odot$ with its radius about 11 km.  The radius for the mass range of $1M_{\odot} ~ - ~ 2 M_{\odot}$  is found to be $  11.2 ~ - ~ 12.2  $km. The calculated deformability parameter for the stiffer EoS employed in this work is in the range  $  4.0 ~ - ~ 0.68$.

What characterizes the approach presented in this work is the stiffening of the EoS due to topology change predicted in the description of  baryonic matter with  skyrmions put on crystal background to access high density. The change is implemented in the properties of the parameters of the effective Lagrangian anchored on chiral-and-scale symmetry of QCD and manifests in nuclear EFT formulated in terms of RG-implemented $V_{lowk}$. Given that the approach describes  fairly well the baryonic matter up to normal nuclear density, it is the changeover of skyrmions to half-skyrmions at a density $\sim (2-3) n_0$ that is distinctive of the model used. This topology change involves no change of symmetries -- and hence no order parameters, therefore it does not belong to the conventional paradigm of phase transitions. But it impacts importantly on physical properties as described in various places in a way that is not present in standard nuclear physics approaches available in the literature.

As has been discussed recently~\cite{hatsuda,baym}, there is another way to produce the stiffening in EoS to access the massive compact stars. It is to implement a smooth changeover from hadronic matter -- more or less well-described -- to strongly correlated quark matter, typically described in NJL model. By tuning the parameters of the quark model so as to produce a changeover at a density $\gsim 2 n_0$, it has been possible to reproduce the features compatible with the properties of observed massive stars.

At first sight, the hybrid hadron-quark model looks quite different from the above  new-BR model. However,  it is not implausible that the two mechanisms share the same physical mechanism. In fact, as argued in \cite{twist}, topology can be traded in for quark degrees of freedom via boundary conditions, i.e., Cheshire cat phenomenon. One can then think of the skyrmion-half-skyrmion transition as depicting baryon-quark transition as one sees in the chiral bag model. In this connection, it would be interesting to see what the hybrid hadron-quark model predicts for the tidal deformation calculated here.

The detections of gravitational wave signals from coalescing binary neutron stars  are expected  to inform  us  on  the tidal deformation\cite{PRD/77/021502}\cite{PRD/81/123016}. Recently  the tidally modified waveforms have been  developed up to  the high frequency of  merger\cite{read} \cite{bernucci}, such that the deformability parameter $\lambda$ , a function of the neutron-star EOS and mass, is measurable within the frequency range of the projected design sensitivity of aLIGO and AdV. It has been also demonstrated in Bayesian analysis that the tidal deformability can be measured to better than $\pm 1 \times 10^{36}$ g cm$^2$ s$^2$ when multiple inspiral events  from three detectors of aLIGO-AdV network~\cite{aLIGO}\cite{aVirgo} are analyzed~\cite{Lackey}. They also show that the neutron star radius can be measured to better than $\pm 1$ km.  Thus  the simultaneous measurement of mass, radius and deformability using gravitational wave detectors could present an exciting possibility to eventually pin down the highly uncertain EoS for the nuclear matter in the mass range of $1M_{\odot} - 2 M_{\odot}$. This would provide a probe for the state of baryonic matter at a density that is theoretically the most uncertain.

The authors would like to thank  Won-Gi Paeng for helpful discussions.  HKL and JL acknowledge the hospitality at APCTP where a part of this work was done. The work was  supported in part by WCU project of Korean Ministry of Education, Science and Technology (Grant No. R33-2008-000-10087-0).

\end{document}